\documentclass[aip,cha,preprint]{revtex4-1}
\usepackage{graphicx}
\usepackage{natbib}
\usepackage{color}

\usepackage{amssymb}
\usepackage{subfigure}
\usepackage{algorithm2e}

\usepackage[latin1]{inputenc}

\begin{document}

\title{Complex network classification using partially self-avoiding deterministic walks}

\author{Wesley Nunes Gon\c{c}alves}
\affiliation{Universidade de S\~{a}o Paulo \\ Instituto de F\'{i}sica de S\~{a}o Carlos (IFSC) \\ wnunes@ursa.ifsc.usp.br}
\author{Alexandre Souto Martinez}
\affiliation{Universidade de S\~{a}o Paulo \\ Faculdade de Filosofia, Ci\^encias e Letras de Ribeir\~ao Preto (FFCLRP)\\ National Institute of Science and Technology in Complex Systems (LNCT-SC) \\ asmartinez@ffclrp.usp.br}
\author{Odemir Martinez Bruno}
\affiliation{Universidade de S\~{a}o Paulo \\ Instituto de F\'{i}sica de S\~{a}o Carlos (IFSC) \\  bruno@ifsc.usp.br}

\begin{abstract}
Complex networks have attracted increasing interest from various fields of science.
It has been demonstrated that each complex network model presents specific topological structures which characterize its connectivity and dynamics.
Complex network classification rely on the use of representative measurements that model topological structures.
Although there are a large number of measurements, most of them are correlated.
To overcome this limitation, this paper presents a new measurement for complex network classification based on partially self-avoiding walks.
We validate the measurement on a data set composed by 40.000 complex networks of four well-known models.
Our results indicate that the proposed measurement improves correct classification of networks compared to the traditional ones.
\end{abstract}



\maketitle


\section{Introduction}

Due to the development of informatics, the acquisition of huge data sets has become possible. 
The analysis of these huge data sets migrated from few individual components (nodes) to a huge number of components all of them interconnected (links) \cite{newman2003}.
It has been noticed that many systems have physically close nodes highly connected and distant nodes are weakly connected. 
As a consequence, a new type of topological structure emerged \cite{watts1998,watts1999,barabasi1999}. 
This new structure interpolates the regular lattice and the random (Erd\~os and R\'enyi) one \cite{erdos1959,solomonoff1951}. 
This change of paradigm has made the junction of the mathematical graph formalism (associated to finite systems) with the statistical physics analysis in a new type of topological structure, which has been named complex network. 
It has been realized that the complex networks can describe several types of topological structures of our daily lives ranging from informatics systems (computer connections in www or internet pages referencing) to biology (protein structures \cite{Alves2007}, metabolic networks) passing through social systems (scientific citation \cite{Batista2006}, actors network, disease and rumor propagation, linguistics \cite{Holanda2004} etc.), and pattern recognition \cite{chalumeau2007,backes2009_2,Goncalves2010,Backes2010prl,}.

To use complex networks formalism in a given system, one must firstly specify which parts of the system form the nodes and how these nodes are interconnected. 
It has been shown that completely different real systems may share a common topological structure. 
Thus, one is concerned in how to differentiate them through topological measurements \cite{backes2009_2}.

For a robust network classification, representative measures must be extracted. 
The problem is how to define a set of measures that is the most appropriate for a specific application.
Several measures have been proposed such as the average number of connections of a node \cite{chalumeau2007,costa2007}, hierarchical degree \cite{costa2006,costa2007}, clustering coefficient \cite{costa2007}, assortativity  etc. 
Nevertheless, many of these measures are correlated, leading to redundancy \cite{costa2007}.
Although  optimal results are not guaranteed, the use of statistical methods (such as principal component analysis or linear discriminant analysis) to select and improve the measure set is an alternative to solve redundancy and subjective selection \cite{costa2007}.

Here we propose to consider networks, with each link having a weight (weighted network) and to use of a new way to classify them. 
We propose to use agents that leave from each node and go to the closest neighbor (the smallest link weight) that has not been visited in the preceding $\mu$ time steps.  
This partially self-avoiding deterministic rule can be modified so that the agent goes to the furthest (the strongest link weight, instead of the closest one) neighbor at each time step.
Although the partially self-avoiding walk rules are simple, the agent trajectories are complicated and lead to an effective exploration of the network.
The closest (furthest) neighbor rule produces trajectories  that explore locally (globally) the network. 
In this study, the partially self-avoiding walks have been applied to classify four kinds of networks: Erd\~os-R\'enyi, geographical, small-world and scale-free. 
This simple procedure allows our method to achieve results in network classification that are better than the traditional ones.
We call attention that a similar technique has been successfully employed in the classification of image texture \cite{backes2009}.

This paper is organized as follows. 
In Sec. \ref{dtw}, a brief review of the results of partially self-avoiding walks is presented. 
In Sec. \ref{method}, we describe the partially self-avoiding walk methods for network classification. 
Numerical experiments and results are presented in Sec. \ref{results}.
Finally, in Sec. \ref{conclusion}, concluding remarks and possible extensions of the method are presented.

\section{Partially Self-avoiding Walk}
\label{dtw}

Random walks in regular or random environment have been extensively studied \cite{metzler2000,derrida1997}.
Nevertheless, deterministic walks in regular \cite{gale1995,bunimovich1992} or random \cite{bunimovich2004} environments also present interesting results.
These results can be applied to a whole variety of practical situations such as: image analysis \cite{campiteli2006,backes2009,backes2011}, pattern recognition \cite{campiteli_2006}, fractal \cite{Blavatska2009}, thesaurus dictionaries \cite{kinouchi2001}, optimization \cite{Xu2000}, etc.

For instance, consider a partially self-avoiding deterministic walk,  where an agent wishes to visit $N$ points randomly distributed in a map of $d$ dimension. 
These points can be considered as sites and the agent can move from one to another following the rule of, at each discrete time step, going to the nearest site not visited in the previous $\mu$ steps. 
The agent performs a partially self-avoiding walk, where the self-avoidance is limited to the memory window $\tau = \mu - 1$.
Although the dynamical deterministic rule is simple, the agent trajectory can be very complicated \cite{lima2001,stanley2001}.  

\begin{figure}[!htbp]
   \begin{center}
     \includegraphics[scale=0.32]{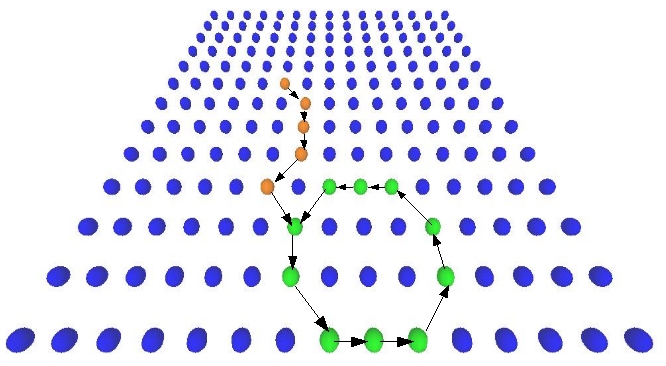}
     \caption{Regular lattice with random weights assigned to the first and second closed neighbors. An agent leaves a given site and moves according to the rule of not returning to the last $\mu$ visited sites. The trajectory is composed of a transient time and attractor are orange and green, respectively.}
     \label{fig:walk}
   \end{center}
\end{figure}

The agent movements depends strictly on the data set configuration and on the starting site \cite{lima2001,stanley2001}. 
They are entirely performed based on a neighborhood table, so that the distances among the sites are simply a way of ranking their neighbors. 
This feature leads to an invariance in scale transformations \cite{campiteli_2006}.
Starting from given point, the trajectory starts with the agent visiting preferentially new points and ends in a cycle, where the same sequence of $p \geq \mu + 1$ points are visited (see Fig. \ref{fig:walk}). 
The beginning of the trajectory has a transient time $t$ and the ending cycle has a period of $p$ or a $p$-attractor. 
Notice that $t$ and/or $p$ are different for different starting points so that a transient time and attractor period joint distribution $S^{(N)}_{\mu,d}(t,p)$ can be computed for the $N$ generated trajectories. 
One obtains the same trajectories for the same point configuration regardless the scale one is dealing. 

The most trivial case to deal with the deterministic agent is to consider $\mu=0$. 
The agent remains in the same site and the trajectory has null transient time and an attractor with period  $p=1$. 
The transient time cycle period joint distribution is simply given by: $S^{(N)}_{0,d}(t,p) = \delta_{t,0} \delta_{p,1}$, where $\delta_{i,j}$ is the  Kronecker delta.
Despite its triviality, this becomes interesting because it is the simplest situation of a  stochastic version of this partially self-avoiding walk \cite{risaugusman2003,martinez:1:2004,Berbert:2010p2922}.

For a memoryless agent ($\mu=1$), at each time step, the agent must leave the current site and go to the nearest one. 
After a very short transient time, the agent becomes trapped by a couple of mutually nearest neighbors. 
The transient time and period joint probability distribution, for $N \gg 1$, can be analytically calculated \cite{tercariol2005}: 
$S_{1,d}^{(\infty)}(t,p) = \Gamma(1+I_d^{- 1})(t+I_d^{- 1}) \delta_{p,2}/\Gamma(t+p+I_d^{- 1})$,
where $\Gamma(z)$ is the gamma function and $I_{d} = I_{1/4} [1/2, (d + 1)/2]$ is the normalized incomplete beta function. In the limit $d \rightarrow \infty$, one is able to calculate it analytically \cite{tercariol:031111}:  
\begin{equation}
S^{(N)}_{2,\infty}(t,p) = \frac{e^{-[3N(t+p-2) (t+p-3)/2]}}{(3 - \delta_{t,0})N} \; .
\label{eq:limit}
\end{equation}


When greater values of $\mu$ are considered, the cycle distribution is no longer peaked at $p_{min} = \mu + 1$, but presents a whole spectrum of cycles with period $p \ge p_{min}$ \cite{lima2001,stanley2001,kinouchi:1:2002,tercariol:031111,tercariol_2007}.

Another possible way to deal with these partially self-avoiding deterministic walks is to consider a two dimensional lattice and randomly distribute random weights to the $(\mu + 1)^{\mbox{th}}$ nearest neighbors so that a given agent, with memory $\mu$, explore the lattice.  
This system is depicted in Fig.~\ref{fig:walk}.

\section{Complex Network Modeling}
\label{method}

Below we describe how this partially self-avoiding walks can be used to classify networks. 
We start presenting how to perform these walks on networks.  
Next we use the transient time cycle period joint distribution to create a vector that characterizes the topology of the network.  

\subsection{Partially Self-avoiding Walks on Networks}
Consider a network represented by a set of  $n$ vertices $V = \{v_{1}, ..., v_{n}\}$ and a set of links $W = \{w_{v_{i},v_{j}} | v_{i}, v_{j} \in V, w_{v_{i},v_{j}} \in \Re \}$, where each component $w_{v_{i},v_{j}}$ represents the link weight that connects vertex $v_i$ to vertex $v_j$. 
Yet, $\eta(v) = \{k \in V | \exists w_{kv}\}$ is a set such that all elements are neighbor of the vertex $v$ and $T = \{t_0, t_1, \dots, t_i | t_j \in V\}$ is a vertex list that store the agent trajectory.

The agent starts the trajectory in a vertex $v_{0}, t_{0} = v_{0}$.
Following the rule of going to the nearest vertex that has not been visited in the previous $\mu$ steps, the agent build his trajectory.
On networks, the movements taken by the agent are completely performed with respect to the set of link weights $W$ and the memory (Eq.\ref{eq:memoria}).
The memory $M_{i}$ is a subset  composed of the last $\mu$ visited vertices of the trajectory $T$
\begin{equation}
 \label{eq:memoria}
 M_i = \{ X \subset T | x = \bigcup_{k=i-\mu}^{i} t_k \} \; .
\end{equation}

Since weight equalities may occur and the agent does not know where to go, one creates the set: 
\begin{equation}
 \zeta = \{v \in V | \arg \min_{v \in \eta(t_i)} w_{v,t_i} | v \notin M_i \}
 \label{eq:walk1}
\end{equation}
to represent a set of vertices, so that these vertices are the closest (with respect to the weights) to the given vertex $t_i$ and do not belong to the memory set $M_i$. 
Depending on the chosen movement  rule, at each time step, the agent leaves its vertex and go to the nearest one (defined by $\arg \min$) or furthest one (switching $\arg \min$ por $\arg \max$ in Eq. \ref{eq:walk1}).
The trajectory is iterated by Eqs. \ref{eq:walk1} and \ref{eq:walk2}. 

To solve possible equalities in $\zeta$, if $\zeta$ has only one element, this is chosen as the following vertex to the agent. 
Otherwise, the equality is solved by a function $\phi$ that returns only a vertex. 
This function may return a vertex randomly chosen or execute a more sophisticated operation
\begin{equation}
 t_{i+1} = \left\{ \begin{array}{ll}
 \varsigma_{1}, & n(\zeta) = 1 \\
 \phi(\zeta), & \textrm{otherwise}\\
 \end{array} \right. \; .
\label{eq:walk2}
\end{equation}

After a transient time $t$, the agent is  trapped in an attractor with period $p$. 
The trajectory can be iterated up to a determined number of steps and it searches for an attractor, or at each time step, determine if an attractor has been reached and finish the trajectory.
The attractor detection is defined by 
\begin{equation}
 \label{eq:atrator}
 \begin{array}{ll}
   \textrm{End} \leftarrow \left\{ \begin{array}{ll}
 \zeta = \emptyset \\
 \exists \sum_{p,it} = 0, & 0 \leq p, it \leq i \\
 \end{array} \right. \\ \\
 \sum_{p,it} = \sum_{j=0}^{p-1} |\varphi(t_{it+j}) - \varphi(t_{it+j+p})| \\
  \end{array}
\end{equation}
considering $\varphi(v)$ as the vertex index $v$, i.e. $\varphi(v_1) = 1, \varphi(v_2) = 2$. 
If $\sum_{p,it}=0$, an attractor with period $p$ and  transient time $t = it-1$ have been detected. 
If $\zeta = \emptyset$, the agent has not found any attractor and $p=0$ with $t=i$.
An efficient computational strategy to find attractors can be found in Ref.~\cite{backes2009}.

\subsection{Signature Vector }

The transient time and cycle period joint distribution $S_{\mu,d}^{(N)}(t,p)$ stores a great quantity of information concerning the partially self-avoiding deterministic walk in a given environment. 
To effectively use these walks, relevant information must be extracted from $S_{\mu,d}^{(N)}(t,p)$. 
This relevant information is stored in a signature vector $\psi_{\mu,din}$, where $din$ represents the dynamics adopted, for instance, the agent   going to the nearest or furthest vertex.

\begin{figure}
   \centering
	 \subfigure[$din = max$]{\includegraphics[width=\textwidth]{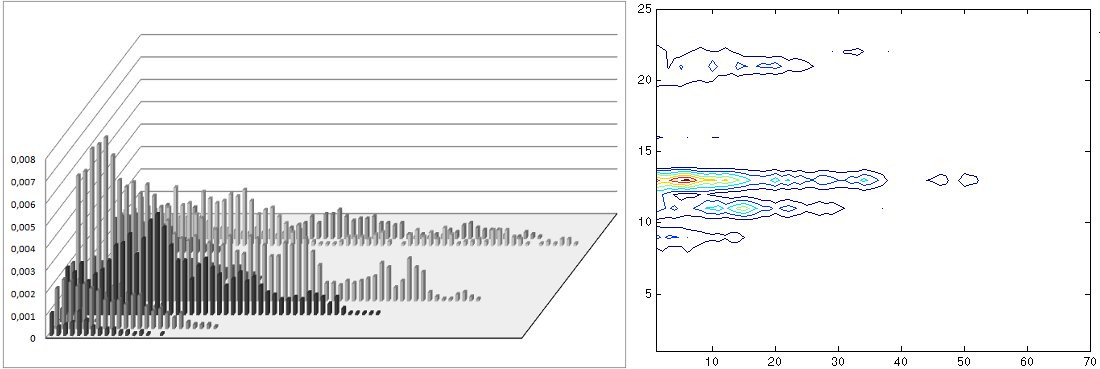}} \\
	 \subfigure[$din = min$]{\includegraphics[width=\textwidth]{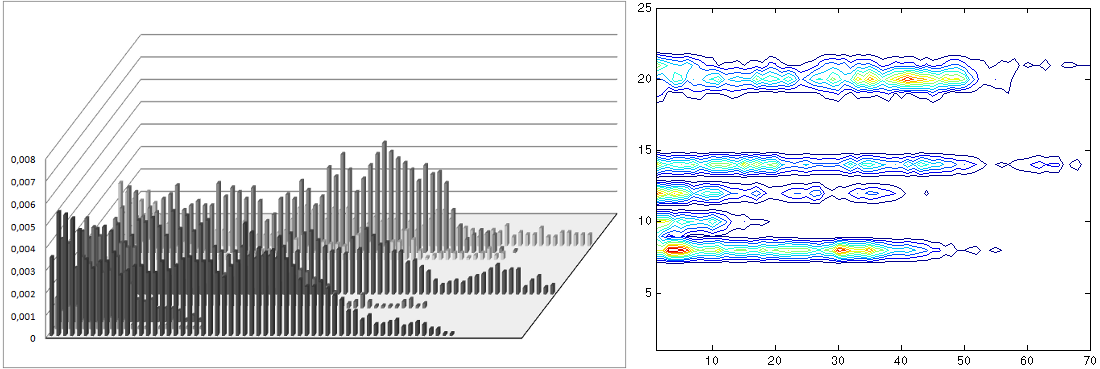}}
   \caption{Transient time $t$ and attractor period $p$ joint distributions obtained from the application of agents with different movement rules on geographical networks with $N = 1000$ and $\left< k \right> = 20$. In a geographical network, the vertices are random and uniformly distributed through a square box and the weight of each edge is proportional to the distance between the vertices. 
   In {\bf (a)}, the walker chooses to go to the closest site, while in {\bf (b)}, the walker goes to the furthest one.}
   \label{fig:minmax}
\end{figure}

An important issue raised about the partially self-avoiding deterministic walk concerns the movement  rule $din$ adopted by the considered agent. 
On one hand, agents guided for the shorter distance are appropriate to find attractors in regions with high homogeneity. 
On the other hand, agents guided for the highest distance find attractors located in regions with low homogeneity.
The use of different movement rules reflect in different joint distributions, allowing the use of information from various sources in the environment characterization \cite{backes2009}.
Trajectories produced by different rules on a graph have distinct patterns for the same graph, as can be seen in Fig.~\ref{fig:minmax}.
In this work, only two movement rules were used: $din=max$, the agent moves to the furthest vertex and $din=min$, the agent moves to the nearest vertex.

The signature vector $\psi_{\mu,din}$ is supposed to characterize the environment where the walks have been performed.
To compose the signature vector, an interesting strategy is to built a histogram $h_{\mu,din}(t+p)$ \cite{backes2009} from the $(t,p)-$joint distribution. 
This histogram represents the number of walks that has a length of $(t+p)$, where $t$ and $p$ are the transient time and period of the attractor, respectively.
From the histogram, $n$ descriptors are used to compose the signature vector $\psi_{\mu,din}$, 
\begin{equation}
 \label{eq:histogram}
 \begin{array}{l}
 \psi_{\mu,din} = [h_{\mu,din}(\mu+1),h_{\mu,din}(\mu+2),...,h_{\mu,din}(t+p),...,h_{\mu,din}(\mu+n)] \\ \\
 h_{\mu,din}(\mu+1) = S^{(N)}_{\mu,2}(0,\mu+1) \\
 h_{\mu,din}(\mu+2) = S^{(N)}_{\mu,2}(0,\mu+2) + S^{(N)}_{\mu,2}(1,\mu+1) \\
 h_{\mu,din}(\mu+3) = S^{(N)}_{\mu,2}(0,\mu+3) + S^{(N)}_{\mu,2}(1,\mu+2) + S^{(N)}_{\mu,2}(2,\mu+1)
 \end{array} \; .
\end{equation}
The first descriptor is on position $\mu+1$, because there is no smaller period.

The $(t,p)-$joint distribution depends on the value of $\mu$ and the movement rule $din$. 
To capture information from different sources and scales, a signature vector $\varphi$ consisting of the concatenation of $\psi_{\mu,din}$ with multiples $\mu$ values and different movement rules $din$ is built:
\begin{equation}
 \label{eq:assinatura}
 \varphi = \left[ \psi_{\mu_1,max}, \psi_{\mu_1,min}, \psi_{\mu_2,max}, \psi_{\mu_2,min}, \dots, \psi_{\mu_M,max}, \psi_{\mu_M,min}\right] \; .
\end{equation}

The algorithm for the proposed measurement is presented below.
First, walks are performed on a complex network $C$ with different memories ($\mu_{1}, ..., \mu_{M}$) and movement rules $din$.
Thus, a joint distribution $S^{(N)}_{\mu_{i}, 2}$ is obtained.
For each joint distribution, a histogram is calculated and a feature vector $\psi_{\mu_{i},din}$ is built.
Finally, the vector of each value of memory and movement rule are concatenated, obtaining a final feature vector with information from various sources and scales.




\section{Analysis of the Proposed Method}
\label{analysis}
\begin{figure}[htbp]
     \centering
     \subfigure[Small world.]{
          \includegraphics[width=1\textwidth]{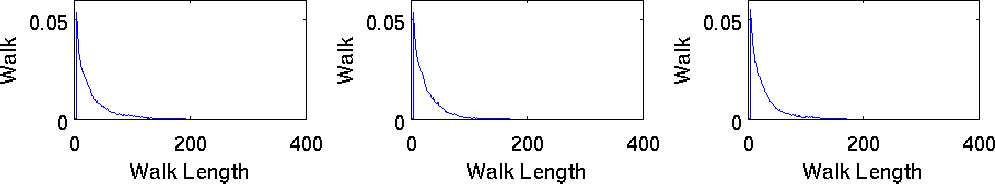}}
     \subfigure[Erd\~os-R\'enyi.]{
          \includegraphics[width=1\textwidth]{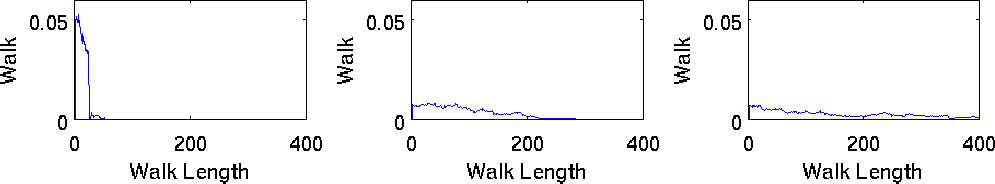}}
     \subfigure[Geographical Network]{
          \includegraphics[width=1\textwidth]{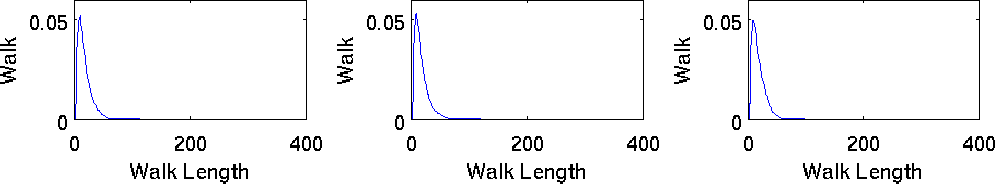}}
     \subfigure[Scale-free.]{
          \includegraphics[width=1\textwidth]{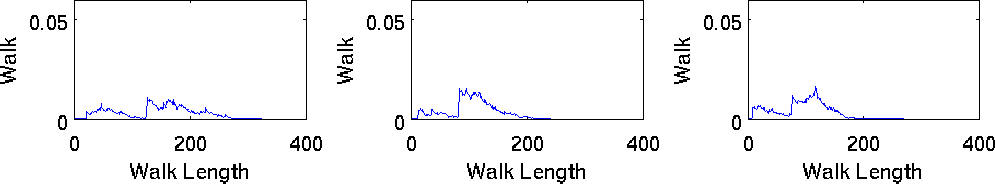}}
     \caption{Histograms of walk length for different complex network models built using $N = 50000$ and vertice degree mean $\left< k \right> = 5$.}
     \label{fig:cnmodels}
\end{figure}

In this section, we present an analysis of the proposed method with regard to the complex network features, such as average degree and number of vertices.
In Fig.~\ref{fig:cnmodels}, histograms of walk length for complex network models built using $N = 50000$ and vertex degree mean $\left< k \right> = 5$ are presented.
For purpose of comparison, each column represents an iteration of the same model and each row represents a complex network model.
It is possible to note that the histograms present distinct patterns and therefore, we can conclude that the walks can be used to characterize each complex network model.
Table \ref{tab:dif_models} shows four statistics calculated from the histograms. From this table, it is shown that the mean of the walk lengths on the scale-free model is the longest.
It occurs because most vertices have few connection, which makes difficult to identify attractors.
On the other hand, the mean of the walk lengths on the geographical networks is the shortest.
In geographical networks, the vertices are connected in proportion to spatial distance, which helps to formation of groups of connected vertices and thus the identification of attractors.

\begin{table}[!htbp]
\centering
	\begin{tabular}{|c|c|c|c|c|}
	\hline
	& \multicolumn{4}{c|}{Statistics of Histograms}\\
	\hline
	\hspace{5mm} Model \hspace{5mm} &  Mean & Standard Deviation & Entropy & Skewness \\
	\hline	
	Small-world & 27.83 & 24.12 & -5.97 & 1.72 \\
	Erd\~os-R\'enyi & 85.57 & 57.42 & -7.66 & 0.64 \\
	Geographical Network & 17.88 & 10.50 & -5.19 & 1.08 \\
	Scale-free & 106.02 & 40.59 & -7.16 & -0.18 \\
	\hline
	\end{tabular}
	\caption{Statistics of the histograms for different complex network models.}
 \label{tab:dif_models}
 \end{table}

\begin{figure}[htbp]
     \centering
     \subfigure[$N = 10000$.]{
          \includegraphics[width=1\textwidth]{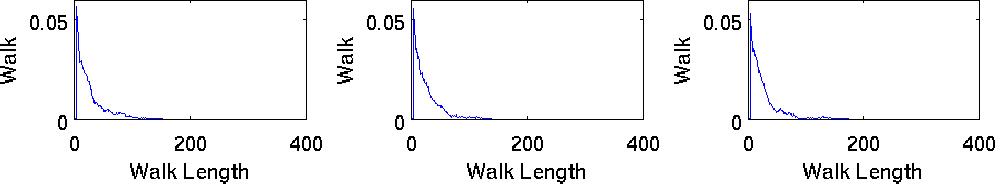}}
     \subfigure[$N = 20000$.]{
          \includegraphics[width=1\textwidth]{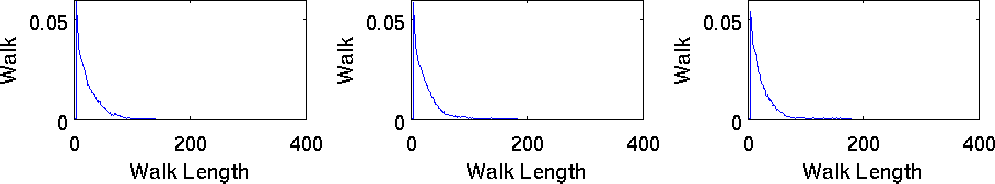}}
     \subfigure[$N = 30000$]{
          \includegraphics[width=1\textwidth]{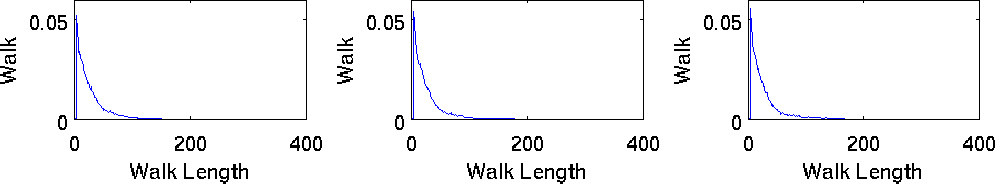}}
     \subfigure[$N = 50000$]{
          \includegraphics[width=1\textwidth]{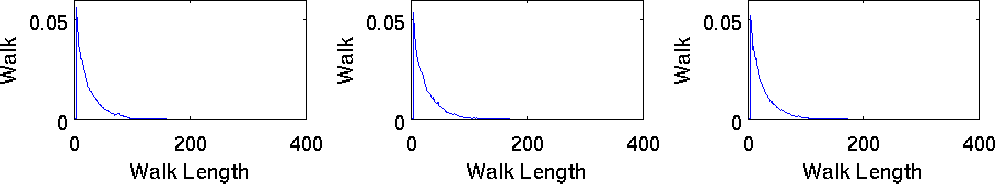}}
     \caption{Histograms of walk length for number of vertices varying from $N = 10000$ to $50000$ on the Small-world model.}
     \label{fig:analysisN}
\end{figure}

Figure \ref{fig:analysisN} illustrates the effect on the histograms for $N$ ranging from $10000$ to $50000$ while keeping $\left< k \right> = 50$ on the small-world model. The histograms are similar as we increase the number of vertices. For the small-world network, the histograms show a peak on walks of small length and decay as the walk length is increased. As an example, the mean, standard deviation, entropy, skewness and kurtosis of the histograms are given in Table \ref{tab:dif_n}.

\begin{table}[!htbp]
\centering
	\begin{tabular}{|c|c|c|c|c|}
	\hline
	& \multicolumn{4}{c|}{Statistics of Histograms}\\
	\hline
	\hspace{5mm} $N$ \hspace{5mm} &  Mean & Standard Deviation & Entropy & Skewness \\
	\hline	
	$10000$ & 27.89 & 23.76 & -5.97 & 1.59\\
	$20000$ & 25.05 & 19.02 & -5.80 & 1.39 \\
	$30000$ & 26.98 & 22.09 & -5.93 & 1.66 \\
	$50000$ & 26.52 & 21.96 & -5.89 & 1.71 \\
	$100000$ & 26.28 & 20.95 & -5.88 & 1.64 \\
	\hline
	\end{tabular}
	\caption{Statistics of the histograms for different values of $N$ on the small-world model.}
 \label{tab:dif_n}
 \end{table}

\begin{figure}[htbp]
     \centering
     \subfigure[$\left< k \right> = 10$.]{
          \includegraphics[width=1\textwidth]{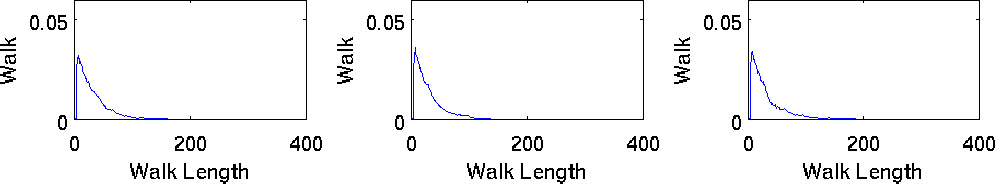}}
     \subfigure[$\left< k \right> = 20$.]{
          \includegraphics[width=1\textwidth]{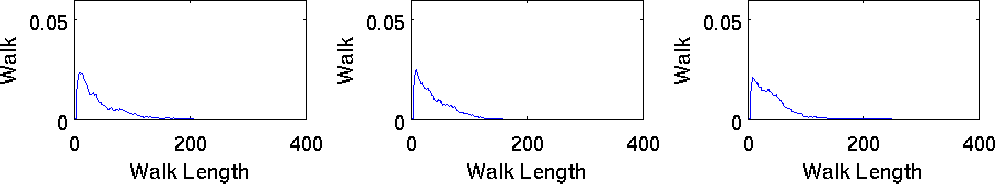}}
     \subfigure[$\left< k \right> = 30$.]{
          \includegraphics[width=1\textwidth]{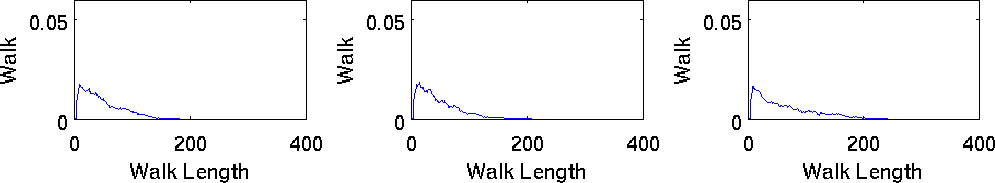}}
     \subfigure[$\left< k \right> = 50$.]{
          \includegraphics[width=1\textwidth]{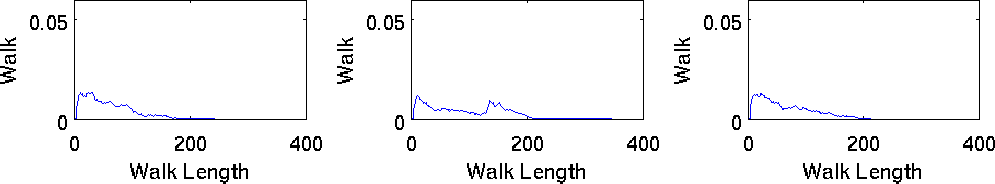}}
     \caption{Histograms of walk length for degree mean varying from $\left< k \right> = 10$ to $50$ on the Small-world model.}
     \label{fig:analysisK}
\end{figure}

The histograms of walk lengths for different values of $\left< k \right>$ are shown in Figure \ref{fig:analysisK}. As we increase the values of $\left< k \right>$, the peak of walks with short length is smoothened.
For high values of $\left< k \right>$, the vertices become more connected and it undermines the traveller in search of an attractor.
Then, walks with longer lengths can be generated more frequently.
Table \ref{tab:dif_k} presents the histogram statistics for different values of $\left< k \right>=20$.

\begin{table}[!htbp]
\centering
	\begin{tabular}{|c|c|c|c|c|}
	\hline
	& \multicolumn{4}{c|}{Statistics of Histograms}\\
	\hline
	\hspace{5mm} $\left< k \right>$ \hspace{5mm} &  Mean & Standard Deviation & Entropy & Skewness \\
	\hline	
	$10$ & 30.86 & 23.11 & -6.14 & 1.35 \\
	$20$ & 41.53 & 31.96 & -6.61 & 1.83 \\
	$30$ & 48.61 & 33.84 & -5.83 & 0.91 \\
	$50$ & 58.01 & 39.68 & -7.07 & 0.76 \\
	\hline
	\end{tabular}
	\caption{Statistics of the histograms for different values of $\left< k \right>$ on the small-world model.}
 \label{tab:dif_k}
 \end{table}

\section{Experiments and Results}
\label{results}

To evaluate partially self-avoiding deterministic walks as a complex network measurement, experiments were performed on a data set composed by 40.000 artificial networks.
The complex network models considered in this data set include: Erd\~os-R\'enyi, small-world, geographical network and scale-free.
The data set consists of artificial complex networks built with random weight for the edges, number of vertices ranging from $N=100$ to $1000$ increasing by steps of $100$, and average degree ranging from $\left< k \right> = 2$ to $20$ increasing by steps of $2$. 
For the geographical network, the vertices were random and uniformly distributed inside a square box.

As in Ref. \cite{backes2009}, $n=4$ descriptors from the histogram have been used to compose the signature vector.  
The statistical analysis reveals that relevant information is concentrated only in the first few elements. 
The signature vector $\varphi$ from each complex network were classified using KNN classifier \cite{mitchell1997} ($k=1$) in a 10-fold cross-validation strategy. 
Since our focus is on modeling, we use a simple classifier rather than a more sophisticated classifier such as support vector machines which have been shown to produce superior results but requires more tuning of parameters.

\subsection{Parameters Evaluation}

\begin{figure}
     \centering
     \subfigure[max.]{
          \includegraphics[width=.48\textwidth]{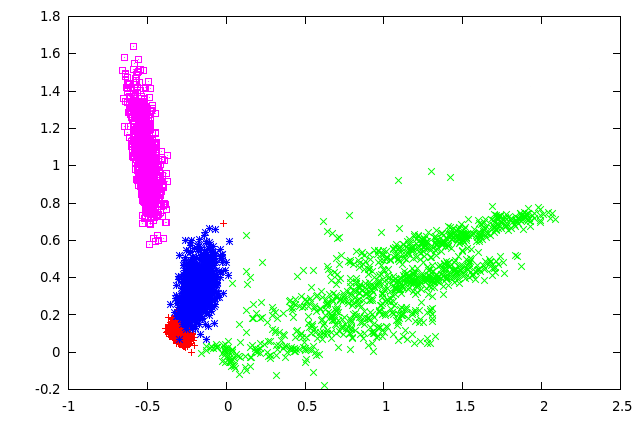}}
     \subfigure[min.]{
          \includegraphics[width=.48\textwidth]{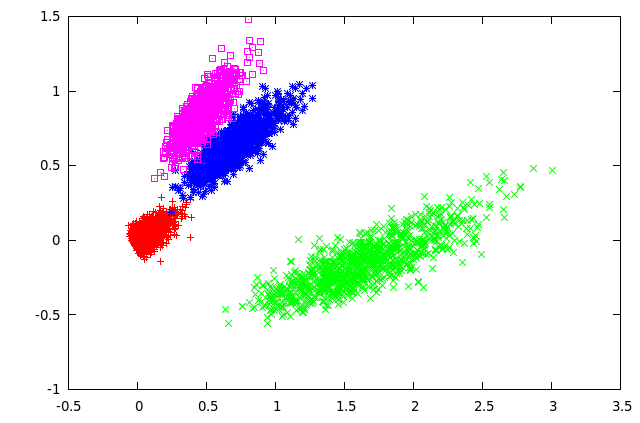}}
     \caption{PCA projection for 4.000 networks obtained by using Erd\~os-R\'enyi, geographical network, small-world and scale-free. Networks were built with $N = 1000$ and $\left< k \right> = 20$ and the walks were performed with $\mu=5$. 
In {\bf (a)}, the walker chooses to go to the closest site, while in {\bf (b)}, the walker goes to the furthest one.    }
     \label{fig:pcaminmax}
\end{figure}

An evaluation of parameters of the partially self-avoiding deterministic walks is presented in the following.
%
Classification results for different values of $\mu$ and movement  rule $din$ are presented in Table \ref{tab:results_brodatz}.
In most cases, movement  rule $min$ achieved better classification results than rule $max$. 
The former found attractors in homogeneous and local regions, i. e. regions where the edge weights are low.
Figure \ref{fig:pcaminmax} shows the Principal Component Analysis (PCA) projection considering two dimensions for both movement rules.
As we can see, the potential of the movement  rule $min$ to obtain separated clusters is evident from this example.
Another important result is that the concatenation of both rules ($[min \: \: max]$) increased the correct classification rate.
This is because the strategy keeps local and global information from the complex network, providing a powerful framework to complex network characterization.

\begin{table}[!hb]
\centering
	\begin{tabular}{|c|c|c|c|c|c|c|}
	\hline
	& \multicolumn{6}{c|}{Memory ($\mu$)}\\
	\hline
	& 0 & 1 & 2 & 3 & 4 & 5 \\	
	\hline	
	$min$ & 74.31 & 87.27 &	73.86 &	64.45 &	61.59 &	58.45 \\
	$max$ & 76.2 &	81.26 &	70.98 &	63.21 &	58.94 &	55.76 \\
	$[min \: \: max]$ & 84.07 & 93.02 & 84.03 & 74.84 & 68.37 & 61.09 \\
	\hline
	\end{tabular}		
\caption{Correct classification rate for $\psi_{\mu,din}$ with different values of $\mu$ and movement  rule $din$.}
\label{tab:results_brodatz}
\end{table}

\begin{figure}[!htpb]
     \centering
     \subfigure[$\mu = 0$.]{
          \includegraphics[width=.48\textwidth]{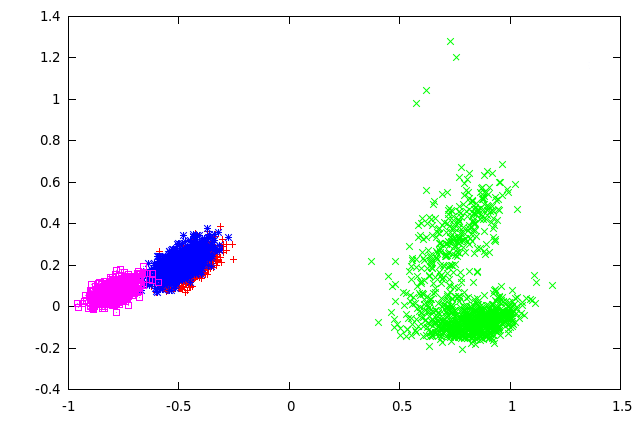}}
     \subfigure[$\mu = 1$.]{
          \includegraphics[width=.48\textwidth]{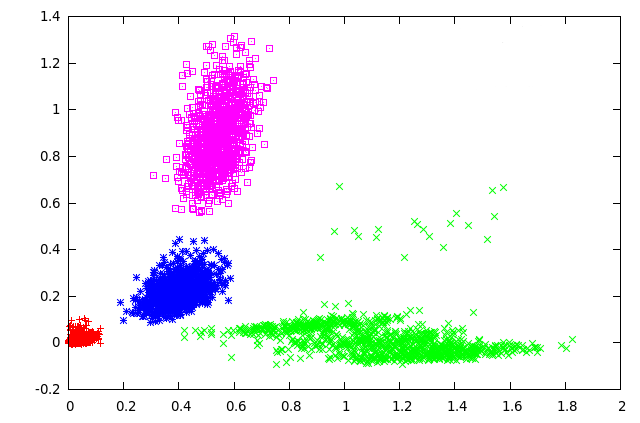}}\\
     \subfigure[$\mu = 2$.]{
          \includegraphics[width=.48\textwidth]{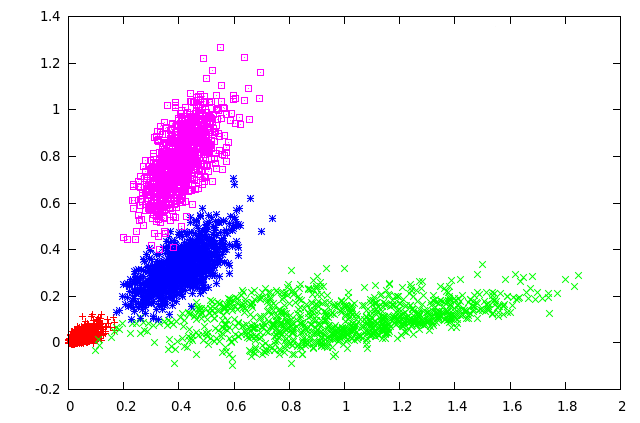}}
     \subfigure[$\mu = 3$.]{
          \includegraphics[width=.48\textwidth]{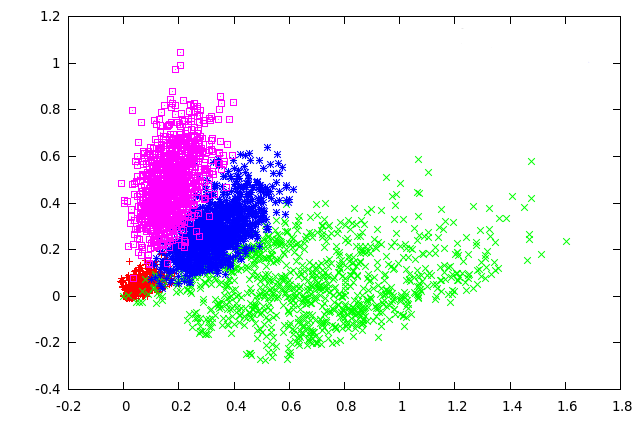}}\\
     \subfigure[$\mu = 4$.]{
          \includegraphics[width=.48\textwidth]{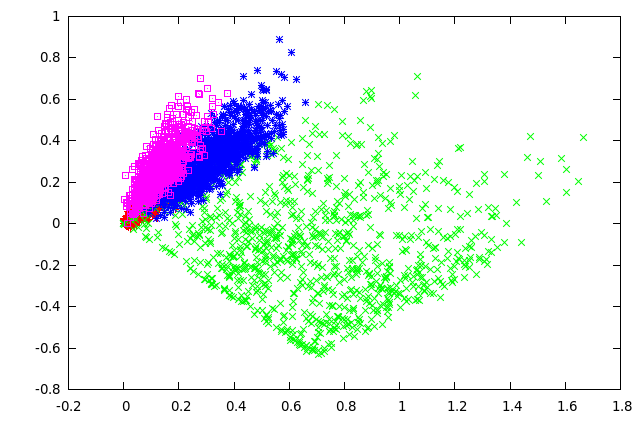}}
     \subfigure[$\mu = 5$.]{
          \includegraphics[width=.48\textwidth]{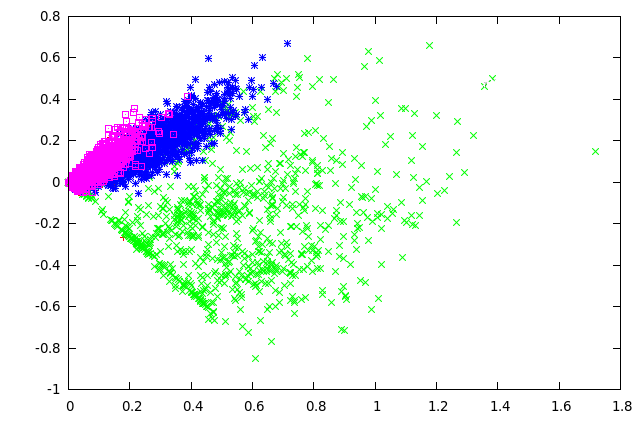}}\\
     \caption{PCA projection for complex network models built with $N = 1000$ and $\left< k \right> = 20$ using deterministic walks with different values of memory and $din=[min, max]$.}
     \label{fig:pcamemorias}
\end{figure}

Results from Table \ref{tab:results_brodatz} also showed that for both movement  rules, the correct classification rate is decreased as the memory is increased, except for $\mu = 0$ which is the trivial case of the deterministic walk.
These results lie on the fact that a walk has more difficulty, as the memory increases, in finding an attractor in the image \cite{backes2009}.
On the other hand, small values of memory $\mu$ perform a better local analysis of the network structure, resulting in an higher correct classification rate.
As an illustration, Figure \ref{fig:pcamemorias} presents the first two PCA discriminant for different values of memory.

\begin{figure}[!htbp]
   \begin{center}
     \includegraphics[width=0.6\textwidth]{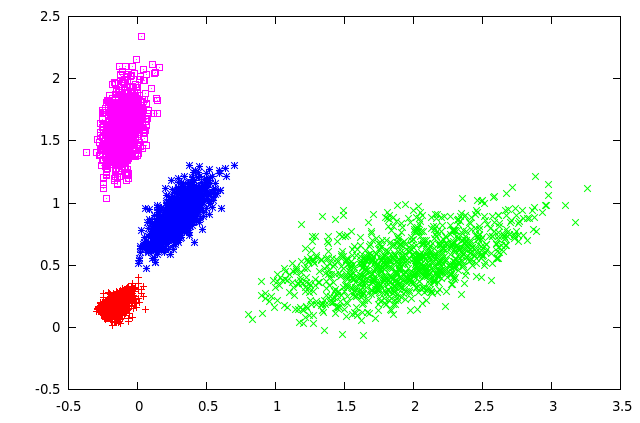}
     \caption{PCA projection of signature vectors composed by the concatenation of memories $0,1,2,3,4$ and $5$. The signatures vectors were extracted from complex networks built with $N = 1000$ and $\left< k \right> = 20$.}
     \label{fig:pcafinal}
   \end{center}
\end{figure}

Results using signature vectors composed by the concatenation of multiple values of memory are presented in Table \ref{tab:multiple_mu}.
This strategy diminishes the importance of individual values of $\mu$ and allows walks with a higher range of lengths, thus providing more robust characterization.
In Figure \ref{fig:pcafinal}, PCA projection of signature vectors composed by multiple values of $\mu$ ($0,1,2,3,4,5$) is shown.

\begin{table}[!hb]
\centering
	\begin{tabular}{|l|c|c|c|}
	\hline
	Memories ($\mu$) & $min$ & $max$ & $[min \: \: max]$ \\	
	\hline	
	$\left\{ 0, 1 \right\}$ & 95.49 & 94.23 & 98.11 \\
	$\left\{ 0, 1, 2 \right\}$ & 95.65 & 94.47 & 98.16 \\
	$\left\{ 0, 1, 2, 3 \right\}$ & 95.73 &	94.68 &	98.19 \\
	$\left\{ 0, 1, 2, 3, 4 \right\}$ & 95.66 & 94.82 & 98.21 \\
	$\left\{ 0, 1, 2, 3, 4, 5 \right\}$ & 95.65 & 94.83 & 98.22 \\
	\hline
	\end{tabular}		
\caption{Correct classification rate for $\varphi$ composed by the concatenation of values $\mu$.}
\label{tab:multiple_mu}
\end{table}

\subsection{Multivariate Analysis of Variance}
\begin{figure}[!htbp]
   \begin{center}
        \subfigure{\includegraphics[width=0.9\textwidth]{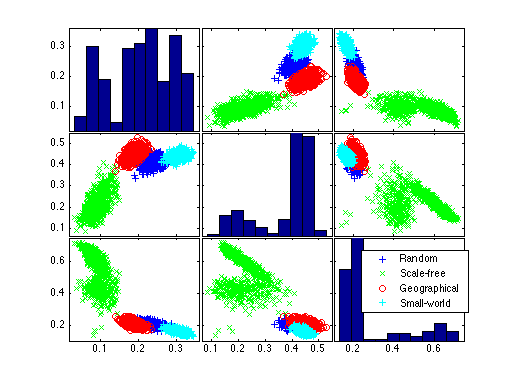}}
        \subfigure{\includegraphics[width=0.48\textwidth]{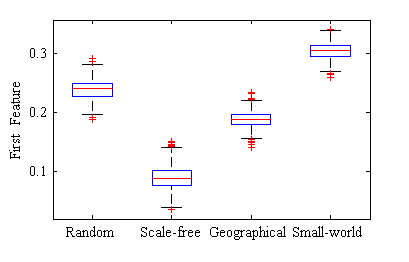}}
        \subfigure{\includegraphics[width=0.48\textwidth]{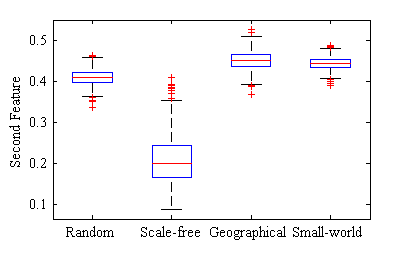}}
     \caption{Scatter plot matrix for the first four features obtained from 4000 complex networks..}
     \label{fig:scatter}
   \end{center}
\end{figure}

Here, we  determine whether the means of the feature vectors differ significantly among the four classes of complex networks.
To answer this question, we apply the multivariate analysis of variance (MANOVA) \cite{Huberty2006}.
The MANOVA takes a set of grouped data composed by the features extracted using the partially self-avoiding walks characteristics.
First, we generate 1000 complex networks for each class (random network, scale-free, geographical and small-world).
For each complex network, from the histogram of the partially self-avoiding walks and extract the feature vector as in the previous experiments.
Thus, one has a data matrix of 4000 rows (1000 samples for each class) and 60 columns (features).
Figure~\ref{fig:scatter} depicts  the scatter plot matrix for the first four features.
On the bottom panel, the figure shows the box plots for different classes.

Considering the means of the four classes by $\mu_{1}, \mu_{2}, \mu_{3}, \mu_{4}$, we perform the MANOVA to test the hypothesis that $\mu_{1} = \mu_{2} = \mu_{3} = \mu_{4}$.
The test of homogeneity (equality) of variance is performed. 
Once the variances are equal, the means are tested and the hypothesis of equal means is rejected even at the 1\% significance level.

\begin{figure}[!htbp]
   \begin{center}
     \includegraphics[width=\textwidth]{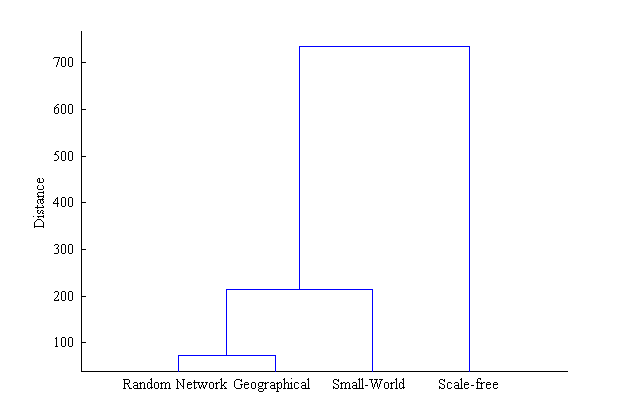}
     \caption{Dendrogram of the class means using the multivariate analysis of variance (MANOVA) takes a set of grouped data composed by the features extracted using the partially self-avoiding walks characteristics of 1000 complex networks for each class (random network, scale-free, geographical and small-world).}
     \label{fig:cluster}
   \end{center}
\end{figure}

In Figure \ref{fig:cluster}, we generate a dendrogram of the class means after the MANOVA.
To obtain the dendrogram, the single linkage method is applied in the matrix of Mahalanobis distances between class means.
The ordenate represents the distance in which two classes is connected.
In this way, one observes that the random network and the geographical network produce features with the most similar characteristics.
Also, the scale-free is the network which produce the most different features from the other three networks.
These results is corroborated by the plots of Figure \ref{fig:scatter}.

\subsection{Correlation with Traditional Measures}

The network classification obtained with the partially self-avoiding  deterministic walks are compared to the traditional methods. 
The results of these comparisons are shown in Table \ref{tab:comparison_random}.
For each traditional measurement for each vertex, the mean value and its standard deviation are estimated. 
Results of in Table \ref{tab:comparison_random} indicate that the classification rate improves with our proposed method from 78.32\% to 98.24\% over the coefficient clustering measurement.
The next highest rate of 71.87\% is obtained for Pearson correlation, where the classification is done with Decision Tree classifier.

\begin{table}[htbp]	
	\centering
		\begin{tabular}{cccc}
			\hline
			Measurement & Knn & Naive Bayes & Decision Tree \\
		  \hline
			Degree & 66.72 & 25.64 & 67.77 \\
			Hier. Degree 2 & 54.67 & 28.45 & 64.47 \\
			Hier. Degree 3 & 54.14 & 32.17 & 64.33 \\
			Weighted Degree & 43.49 & 25.63 & 54.72 \\
			Weighted Hier. Degree 2 & 51.98 & 28.50 & 62.61 \\
			Weighted Hier. Degree 3 & 53.27 & 32.11 & 63.51 \\
			Clustering Coefficient & 68.13 & 69.19 & 78.32 \\
			Pearson Correlation & 67.02 & 58.96 & 71.87 \\
			Proposed Measurement & 98.22 & 80.12 & 98.24 \\
			\hline			
		\end{tabular}
	\caption{Comparison between measurements extracted from the complex networks.}
	\label{tab:comparison_random}
\end{table}

An interesting strategy for classifying complex networks involves the concatenation of signature vectors extracted by different measurements.
In this work, we considered two concatenations: a) concatenation of all traditional measurements and b) concatenation of the proposed measurement and all traditional ones.
Table \ref{tab:comparison} presents the experimental results obtained by the combinations on the complex network model classification. 
When all traditional measurements are concatenated, the result obtained only by the proposed measurement is still equivalent (99.18\% against 98.22\% using Knn classifier and 99.10\% against 98.24\% using Decision Tree classifier). 
The combination between traditional measurement and the proposed one achieved the highest correct classification rate of 99.96\%.

In principle, signature vector composed by more measurements can provide a better complex network classification.
This fact suggests that each network model has specific topological features modeled by different traditional measurements.
Experimental results obtained by the proposed measurement suggest that this new measurement can model most of the topological features of the complex network models.
The proposed method results are even more important since the concatenation of an excessive number of measurements may affect the quality of complex network classification \cite{costa2007}.

\begin{table}[htbp]	
	\centering
		\begin{tabular}{cccc}
			\hline
			Measurement & Knn & Naive Bayes & Decision Tree \\	 
			\hline									
			Traditional Measurements & 99.18 & 80.96 & 99.10 \\			
			Proposed Measurement & 98.22 & 80.12 & 98.24 \\
			Traditional and Proposed Measurement & 99.96 & 81.54 & 99.92\\
			\hline			
		\end{tabular}
	\caption{Comparison between combination of measurements.}
	\label{tab:comparison}
\end{table}

\section{Conclusion}
\label{conclusion}

Here, we have presented a new method to classify complex network based on self-avoiding deterministic trajectories generated by agents leaving from each site of a given weighted network.  
The agent movements are given by a rule which may characteristics is to forbid visitation to vertex recently visited, with a memory $\mu$. 
The trajectories formed by these agents, have a transient time and finish in a cycle with an attractor with a given period. 
Form the transient time and attractor period joint distribution, we build a signature vector. 
It is with this vector that one classifies the networks types.
So that, the agents explore and characterize the complex network topology.
Combining walks performed with different values of $\mu$ and $din$, it is possible to include information from different sources and scales and then improve the complex network characterization.

Promising results have been obtained on a data set composed by 40.000 networks of 4 well-known models.
Experimental results indicate that the proposed method improves recognition of models compared to the traditional measurements.
In addition, our method makes the modeling of complex network feasible and simple, since only the use of this new measurement is enough to obtain satisfactory results of classification.

\section*{Acknowledgements}
W.N.G. acknowledges support from FAPESP (2008/03253-9).
A. S. M. acknowledges the Brazilian agency CNPq (305738/2010-0 and 476722/2010-1) for support. 
O.M.B. acknowledges support from CNPq (Grant \#308449/2010-0 and \#473893/2010-0) and FAPESP (Grant \# 2011/01523-1). 


%

\end{document}